\documentclass[conference]{IEEEtran}
\IEEEoverridecommandlockouts

\usepackage{cite}
\usepackage{amsmath,amssymb,amsfonts}
\usepackage{amsthm}
\usepackage{booktabs}
\usepackage{multirow}
\usepackage{mathrsfs}
\usepackage{hhline}
\usepackage{graphicx}
\usepackage{textcomp}
\usepackage{xcolor}
\usepackage{bm}
\usepackage{algorithm}      
\usepackage{algpseudocode}  
\usepackage[table]{xcolor}

\usepackage{hyperref}
\hypersetup{
    colorlinks=false
}

\algtext*{EndIf}
\algtext*{EndFor}

\theoremstyle{definition}

\def\BibTeX{{\rm B\kern-.05em{\sc i\kern-.025em b}\kern-.08em
    T\kern-.1667em\lower.7ex\hbox{E}\kern-.125emX}}
\begin{document}

\title{GapAttack: Beyond Target Alignment in Multi-Modal Adversarial Attacks}

\author{\IEEEauthorblockN{Maitreya Sawai}
\IEEEauthorblockA{
\textit{Louisiana State University}\\
Baton Rouge, LA \\
msawai1@lsu.edu}
\and
\IEEEauthorblockN{Xugui Zhou}
\IEEEauthorblockA{
\textit{Louisiana State University}\\
Baton Rouge, LA \\
xuguizhou@lsu.edu}
}

\maketitle

\begin{abstract}

Multi-modal encoders expose diverse downstream tasks to a shared adversarial attack surface. Existing embedding attacks emphasize target alignment, yet we show that stronger alignment does not necessarily improve downstream attack success. We investigate the role of modality geometry and propose \textsc{GapAttack}, which jointly optimizes target alignment and adversarial embedding placement along a modality axis defined by image and text embedding centroids. Across multiple datasets, encoders, and downstream tasks, \textsc{GapAttack} outperforms baselines by up to 9.2 percentage points in caption generation while maintaining near-perfect targeted classification success. 
These findings highlight the importance of accounting for modality geometry when attacking shared multi-modal representations.

\end{abstract}

\section{Introduction}
\label{sec:intro}

Multi-modal encoders such as CLIP~\cite{radford2021clip} and ImageBind~\cite{girdhar2023imagebind} are deployed as shared representation backbones, mapping inputs from different modalities into a common embedding space. One encoder is trained once and reused across diverse downstream tasks like retrieval~\cite{radford2021clip}, classification~\cite{zhou2022coop}, generation~\cite{ramesh2022hierarchicaltextconditionalimagegeneration, mokady2021clipcapclipprefiximage}, and vision-language applications~\cite{liu2023llava} that can operate independently of one another~\cite{you2022learningvisualrepresentationmodalityshared}. Sharing an encoder across downstream applications can improve resource efficiency while creating a common attack surface at the embedding interface.

An adversary can perturb an input to align its embedding with a chosen target, potentially compromising multiple downstream components, including those unknown at attack time. However, these effects are not uniform across consumers, since a perturbation that reliably induces misclassification may fail to achieve the adversary's objective in captioning or generation \cite{chen2025revisiting,zhou2025runtime,zhou2022robustness}. Consequently, evaluating the encoder or a single consumer is insufficient to characterize the attack’s impact across downstream applications.

Prior work has advanced targeted attacks on shared multi-modal
embeddings through target alignment, same-modal target construction,
and proxy-based optimization.
{Alignment-based attacks} craft adversarial examples (AEs) by perturbing an input until its embedding maximizes cosine similarity to an attacker-chosen target in the shared embedding
space~\cite{zhang2024illusions,
inkawhich2023adversarialattacksfoundationalvision}.
CrossFire~\cite{dou2024adversarialattacksmultimodalmodels} addresses cross-modal mismatch by converting the target into the source modality before aligning the adversarial embedding with the converted target.
PTA~\cite{zhang2026improvinggeneralizabilityundetectabilitytargeted} incorporates both target-modal and source-modal proxies to improve generalizability and undetectability.
Related studies also report task-dependent attack effectiveness,
highlighting that success on one downstream consumer does not
necessarily imply success on
others~\cite{zhao2023evaluatingadversarialrobustnesslarge,
liu2025bridgingtaskgapmultitask,
silva2025attackingattentionfoundationmodels,
daw2024hidinginplainsighthipsattackclip}.

Despite these advances, it remains unclear how embedding placement relative to modality regions affects downstream attack success beyond target alignment.
Studies of contrastive multi-modal encoders show that different modalities occupy distinct regions of the shared embedding space, with their separation referred to as the \emph{modality gap}~\cite{liang2022mindgapunderstandingmodality}.
Prior work has related this geometry to robustness under embedding perturbations~\cite{chowers2026modalitygapbugfeature}.
Target similarity alone does not generally determine an adversarial embedding's position relative to these regions.
This motivates our central question:
\emph{At comparable target-alignment levels, how does an adversarial embedding's placement relative to modality regions affect downstream attack success, particularly for consumers that generate captions or images?}

\begin{figure*}[t]
  \centering
  \includegraphics[width=\linewidth]{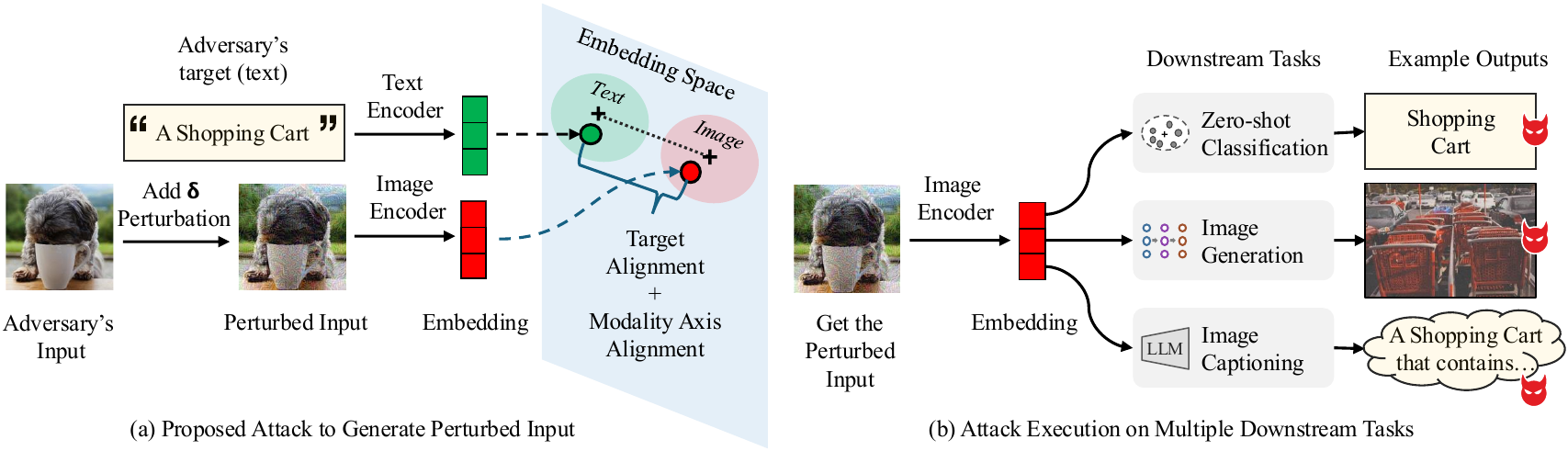}  \caption{\textbf{Illustration of \textsc{GapAttack}}: A single perturbed input can deceive multiple downstream consumers at once.}
  \label{fig:overview}
  \vspace{-1em}
\end{figure*}

We address this question by analyzing adversarial embedding placement
along the \emph{modality axis}, defined by the direction connecting the two modality centroids. 
Our results show that stronger target alignment does not necessarily
improve downstream attack success, and that controlling placement
along this axis can improve success even at matched alignment levels.
Based on these findings, we propose \textsc{GapAttack} (Fig.~\ref{fig:overview}), which jointly optimizes target alignment and embedding placement along the \emph{modality axis}. 
We evaluate the attack using OpenCLIP ViT-B/32 and ImageBind across multiple datasets and downstream tasks, including zero-shot classification, retrieval, captioning, and image generation.

The {main contributions} are summarized as follows:
\begin{enumerate}\itemsep2pt
  \item We analyze adversarial embedding placement along the modality axis and show that it influences attack success on the evaluated generative consumers even at matched target-alignment levels. 
  \item We propose \textsc{GapAttack}, which jointly optimizes target alignment and adversarial embedding placement along a centroid-defined modality axis.
  \item We evaluate \textsc{GapAttack} across multiple datasets, encoders, and downstream tasks.
  It improves captioning attack success by up to 9.2 percentage points over the compared baselines while maintaining near-perfect targeted classification success. Matched-alignment experiments on ImageBind further show improvements in  captioning and image generation.
  \end{enumerate}

\section{Related Work}
\label{sec:related}

\subsection{Attacks on Shared Multi-modal Embeddings}
Attacks on shared multi-modal representations exploit several distinct mechanisms (see Table~\ref{tab:related}).
Adversarial Illusions uses embedding alignment, a central property of contrastive representations~\cite{wang2022understandingcontrastiverepresentationlearning}, to steer perturbed inputs toward cross-modal targets without requiring knowledge of downstream consumers~\cite{zhang2024illusions}. 
Other attacks exploit hubness to make injected items relevant to many unrelated queries~\cite{zhang2026hubness}, use adversarial images as soft prompts to influence a vision-language model’s responses~\cite{zhang2025selfinterpreting}, or jailbreak multimodal pipelines~\cite{shayegani2024plugandpray}. 
These attacks differ in their objectives and access assumptions.

Prior work examines downstream attack behavior through surrogate-based transfer~\cite{zhao2023evaluatingadversarialrobustnesslarge}, gradient conflicts in untargeted cross-task attacks~\cite{liu2025bridgingtaskgapmultitask}, attention- versus embedding-based perturbations~\cite{silva2025attackingattentionfoundationmodels}, and loss balancing for targeted object removal from captions~\cite{daw2024hidinginplainsighthipsattackclip}. 
For targeted embedding attacks, CrossFire converts the target into the source modality before embedding alignment~\cite{dou2024adversarialattacksmultimodalmodels}, while PTA uses target-modal and source-modal proxies to improve generalizability and undetectability, respectively~\cite{zhang2026improvinggeneralizabilityundetectabilitytargeted}.
Our study complements these approaches by analyzing how placement along a centroid-defined modality axis affects downstream attack success at matched target-alignment levels. \textsc{GapAttack} incorporates an explicit term controlling the adversarial embedding's projection onto this axis.

\begin{table}[t]
\caption{Relation to prior work on Embedding Attacks and Geometry.}
\vspace{-2em}
\begin{center}
\resizebox{\columnwidth}{!}{%
\setlength{\tabcolsep}{1pt}
\begin{tabular}{|l|c|c|c|c|c|}
\hline
\rowcolor[gray]{0.88}
\textbf{Work} & \textbf{Att.} & \textbf{Def.} & \textbf{Embedding}
 & \textbf{Cross-modal} & \textbf{Unknown Gen.} \\
\rowcolor[gray]{0.88}
 & & & \textbf{Geometry} & \textbf{Target} & \textbf{consumer} \\
\hline
Illusions~\cite{zhang2024illusions}        & $\checkmark$ & $\times$ & $\times$ & $\checkmark$ & $\checkmark$ \\
\cline{1-6}
Hubness~\cite{zhang2026hubness}            & $\checkmark$ & $\times$ & $\times$ & $\checkmark$  & $\times$ \\
\cline{1-6}
AttackVLM~\cite{zhao2023evaluatingadversarialrobustnesslarge} & $\checkmark$ & $\times$ & $\times$ & $\checkmark$  & $\checkmark$ \\
\cline{1-6}
MT-AdvCLIP~\cite{liu2025bridgingtaskgapmultitask} & $\checkmark$ & $\times$ & $\times$ & $\times$  & $\times$ \\
\cline{1-6}
PTA~\cite{zhang2026improvinggeneralizabilityundetectabilitytargeted} & $\checkmark$ & $\times$ & $\times$ & $\checkmark$ & $\times$ \\
\cline{1-6}
HiPS~\cite{daw2024hidinginplainsighthipsattackclip}, Attn.~\cite{silva2025attackingattentionfoundationmodels}  & $\checkmark$ & $\times$ & $\times$ & $-$  & $\checkmark$ \\
\cline{1-6}
CrossFire~\cite{dou2024adversarialattacksmultimodalmodels} & $\checkmark$ & $\times$ & $\times$ & $\checkmark$  & $\checkmark$ \\
\hline
Mind the Gap~\cite{liang2022mindgapunderstandingmodality} & $\times$ & $\times$ & $\checkmark$ & $-$  & $-$ \\
\cline{1-6}
Gap/robust.~\cite{chowers2026modalitygapbugfeature} & $\times$ & $\times$ & $\checkmark$ & $-$  & $-$ \\
\hline
Consensus~\cite{akbarian2026breakingillusionconsensusbasedgenerative}, A$^4$D~\cite{krakover2026a4d} & $\times$ & $\checkmark$ & $\times$ & $-$ & $\checkmark$ \\
\cline{1-6}
CLIPure~\cite{zhang2025clipure} & $\times$ & $\checkmark$ & $\times$ & $-$ & $\times$ \\
\cline{1-6}
COLA~\cite{zhu2025cola} & $\times$ & $\checkmark$ & $\checkmark$ & $-$ & $\times$ \\
\hline
\rowcolor[gray]{0.95}
\textbf{Ours} & $\checkmark$ & $\times$ & $\checkmark$ & $\checkmark$  & $\checkmark$\\
\hline
\end{tabular}
\label{tab:related}
}
\end{center}
\vspace{-2em}
\end{table}

\subsection{Modality Gap and Robustness} 
Prior work investigates both the structure of modality gaps and their implications for downstream learning. 
Studies characterize the separation of modality distributions~\cite{liang2022mindgapunderstandingmodality}, reduce the gap to improve cross-modal representations~\cite{fahim2024itsmodalitygapcharacterizing, eslami2025mitigate,grassucci2025closing}, preserve and compensate for it in continual learning~\cite{Huang_2025_ICCV}, and correct modality mismatch in text-only image captioning~\cite{fonseca2025textonlytrainingimagecaptioning}.

Related robustness research connects modality geometry to sensitivity under embedding perturbations~\cite{chowers2026modalitygapbugfeature}, while Robust CLIP and LORE improve encoder robustness through unsupervised adversarial fine-tuning that keeps perturbed-input embeddings close to clean reference embeddings from the original encoder~\cite{schlarmann2024robustclip,khodabandeh2025lore}.
Our work instead examines targeted input perturbations against a fixed encoder, focusing on how adversarial embedding placement within its existing modality geometry influences downstream attack success.

\section{Methodology}
\label{sec:residue}

\subsection{Modality Gap in Multi-modal Embeddings}
\label{subsec:what-alignment}

Let $\theta = \{\theta^{(i)}\}_{i\in\mathscr{M}}$ denote a multi-modal encoder that maps inputs into a shared embedding space. We use unit-normalized embeddings and write $\langle a,b\rangle$ for the inner product, which equals cosine similarity for unit-norm vectors.
Embeddings from different modalities  can concentrate in distinct regions of this space, a phenomenon known as the
\emph{modality gap}~\cite{liang2022mindgapunderstandingmodality}. 

Consider distinct modalities $m,\bar m\in\mathscr{M}$ with
inputs $x_m$ and $y_{\bar{m}}$, respectively. 
We define the centroid-difference vector
\begin{equation}
  g = \mathbb{E}\big[\theta^{(\bar m)}(y_{\bar m})\big] - \mathbb{E}\big[\theta^{(m)}(x_m)\big],
  \label{eq:axis}
\end{equation}
where the expectations are taken over the respective modality input distributions and estimated from clean reference samples. 
For $g\ne0$, the unit vector $\hat g=g/\|g\|_2$ defines the \emph{modality axis}, pointing from the centroid of modality $m$ toward that of modality $\bar m$.
For any two embedding vectors, $u$ and $v$, we define their \emph{signed separation along the modality axis} as $\mathscr{G}(u,v) \;=\; \big\langle v-u,\ \hat g \big\rangle$.
A positive value indicates that $v$ has a larger projection
onto $\hat{g}$ than $u$. 
We incorporate this signed separation into the attack objective to control adversarial embedding placement along the modality axis.

\subsection{Threat Model}
\label{sec:threat}
We consider a two-stage machine learning pipeline, in which a fixed multi-modal encoder maps an input to an embedding, and downstream consumers use that embedding to produce predictions or generate content.

\textbf{Adversary's Goal.}
Given an input $x$ and an adversary-chosen target $y$ in a modality supported by the encoder, the adversary aims to construct a perturbed input $x'=x+\delta$ that steers downstream outputs toward the semantics of $y$ while preserving the perceived semantic content of $x$.

\textbf{Adversary's Capability.}
The adversary can perturb only the input, subject to $\|\delta\|_{\infty}\le\epsilon$. 
It has no access to original training datasets of the
encoder or downstream consumers.
During attack construction, downstream consumer identities, parameters, and outputs are unavailable to the adversary.

\textbf{Adversary's Knowledge.}
Following prior embedding attacks~\cite{zhang2024illusions,zhang2026hubness}, we assume white-box access to the fixed encoder $\theta$, including its parameters and input gradients. 
The adversary estimates the modality centroids and the axis $\hat g$ in Eq.~\eqref{eq:axis} from available clean reference samples of the relevant modalities, and reuses these estimates across attacks on the same encoder.

\subsection{Limitations of Target Alignment}
\label{subsec:not-sufficient} 
\begin{figure}[t]
    \centering
    \includegraphics[width=\columnwidth]{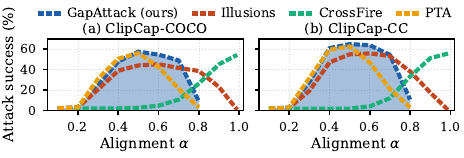}
    \vspace{-2em}

    \caption{\textbf{Example:} Downstream success is not monotonic in target alignment}
    \label{fig:alpha_curves}
    \vspace{-1em}
\end{figure}

Higher target alignment does not necessarily yield higher downstream attack success. We denote by $\alpha$ the cosine similarity between the perturbed-input embedding and the target embedding.
For each evaluated attack, we record the embedding when its target alignment first reaches each prescribed threshold, and evaluate captioning attack success using two ClipCap captioners~\cite{mokady2021clipcapclipprefiximage}:
ClipCap-COCO and ClipCap-CC, trained on MS-COCO \cite{10.1007/978-3-319-10602-1_48} and Conceptual Captions, respectively.
As shown in Fig.~\ref{fig:alpha_curves}, attack success is not monotonic in target alignment.
This observation motivates examining whether explicit control of modality-axis placement can improve downstream attack success at comparable target-alignment levels.

\subsection{GapAttack}
\label{sec:attack}

An alignment-only objective does not explicitly control adversarial embedding placement along the modality axis.
Motivated by the preceding observations, \textsc{GapAttack} augments target alignment with the signed axial separation $\mathscr{G}$ defined above:
\begin{equation}
\label{eq:GapAttack}
\begin{aligned}
\underset{\|\delta\|_{\infty}\le\epsilon}{\operatorname{maximize}}
\quad
& \underbrace{
  \big\langle \theta^{(m)}(x_m+\delta),
  \theta^{(\bar{m})}(y_{\bar{m}}) \big\rangle
}_{\text{target alignment } \alpha} \\
& {}+ \lambda\,
\underbrace{
  \mathscr{G}\big(
    \theta^{(m)}(x_m+\delta),
    \theta^{(\bar{m})}(y_{\bar{m}})
  \big)
}_{\text{signed separation along modality axis}} 
\end{aligned}
\end{equation}
The first term increases similarity to the target embedding. For a fixed target, maximizing the second term with $\lambda>0$ favors smaller projections of the adversarial embedding onto $\hat g$.
Since $\hat g$ points from the source-modality centroid toward the target-modality centroid, this term biases placement toward the source side along the modality axis.
The weight $\lambda$ controls the trade-off between target alignment and axial placement; $\lambda=0$ recovers the alignment-only objective.

\textbf{Encoder-based calibration.}
On the evaluated encoders, clean image embeddings concentrate in a narrow band along the modality axis.
We therefore calibrate $\lambda$ by matching the mean axial projection of adversarial embeddings to that of clean inputs, using only encoder outputs.
Let $\delta_{\lambda,\alpha}$ denote the perturbation recorded when target alignment first reaches a prescribed threshold $\alpha$ under weight $\lambda$.
For calibration, we fix $\alpha=0.5$ and seek
\begin{equation}
\mathbb{E}_{x_m}\big[\big\langle \theta^{(m)}(x_m+\delta_{\lambda,\alpha}),\, \hat g \big\rangle\big]
\;=\;
\mathbb{E}_{x_m}\big[\big\langle \theta^{(m)}(x_m),\, \hat g \big\rangle\big],
\label{eq:anchor}
\end{equation}
The expectations in Eq.~\eqref{eq:anchor} are estimated over the inputs to be attacked: the left-hand side uses their adversarial embeddings at alignment $\alpha$, and the right-hand side uses their clean embeddings.
The same inputs and attack targets are retained across candidate $\lambda$ values.
We evaluate the left-hand side for different $\lambda$ values, identify two values whose mean projections bracket the clean-input mean, and interpolate to estimate $\lambda$.
This procedure yields $\lambda=0.5$ for OpenCLIP and $\lambda=0.40$ for ImageBind. The resulting weight is reused across attacks on the same encoder. This calibration matches the mean axial projection without using downstream consumer parameters or outputs.

\textbf{Attack generation.}
With $\hat g$ and $\lambda$ fixed, we optimize Eq.~\eqref{eq:GapAttack} using projected gradient descent~\cite{madry2018towards}. For evaluations at prescribed alignment levels, we retain the first iterate reaching each threshold, as described in Sec.~\ref{subsec:not-sufficient}.
Algorithm~\ref{alg:attack_pipeline} summarizes the attack procedure, with the modality axis and weight calibrated before attack generation.
The effect of embedding placement depends on how downstream consumers use the shared representation.

\begin{algorithm}[h]
\caption{\textsc{GapAttack} at a prescribed alignment level}
\label{alg:attack_pipeline}
\begin{algorithmic}[1]
\Require Input $x_m$, target $y_{\bar m}$, encoder $\theta$
\Require Precomputed $\hat g,\lambda$; perturbation budget
         $\epsilon$; alignment threshold $\alpha$
\Ensure First input reaching $\alpha$, or an unreached-threshold flag

\State $v \gets \theta^{(\bar m)}(y_{\bar m})$
\State Generate PGD iterates $\{\delta_t\}_{t=0}^{T}$
       for Eq.~\eqref{eq:GapAttack} using the configuration
       specified in Experimental Setup
\For{$t=0,\ldots,T$}
    \If{$\langle\theta^{(m)}(x_m+\delta_t),v\rangle
         \ge\alpha$}
        \State \Return $x_m+\delta_t$
    \EndIf
\EndFor
\State \Return \texttt{threshold not reached}
\end{algorithmic}
\end{algorithm}
\textit{Margin consumers.}
Zero-shot classification and retrieval rank a fixed candidate set $\{c_1,\dots,c_C\}$ by embedding similarity. The candidates are embedded class prompts for classification and embedded gallery items for retrieval. 
For targeted top-1 classification or retrieval, the attack succeeds when the target candidate $c_t$ scores higher than every other candidate:
\begin{equation}
  \big\langle \theta^{(m)}(x_m+\delta), c_t\big\rangle \;>\; \max_{j\ne t}
\big\langle \theta^{(m)}(x_m+\delta), c_j\big\rangle .
  \label{eq:margin}
\end{equation}

\textit{Generative consumers.}
Captioners and image generators produce outputs through learned decoders conditioned on the embedding.
For the evaluated generative consumers, target alignment alone does not fully characterize downstream attack success; placement along the modality axis also matters.
We analyze this dependence in Sec.~\ref{sec:experiments}.

\begin{table*}[!t]
\caption{Performance Comparison with Baselines (\%) on OpenCLIP Vit at each attack's best alignment $\bm{\alpha^{\star}}$.}
\begin{center}
\small
\setlength{\tabcolsep}{0.7pt}
\begin{tabular}{l||c|c|c||c|c||c|c}
\hline
\rowcolor[gray]{0.75}
 & \multicolumn{3}{c||}{ImageNet-V2} & \multicolumn{2}{c||}{MS-COCO} & \multicolumn{2}{c}{STL-10} \\
\hhline{-||-|-|-||-|-||-|-}
Attack & Zero-shot & COCO Cap. Gen. & CC Cap. Gen. & Zero-shot & COCO Cap. Gen. & Zero-shot & COCO Cap. Gen. \\
\hhline{=#=|=|=#=|=#=|=}
Illusions~\cite{zhang2024illusions} & 100.0 ($\bm{\alpha^{\star}}$=0.50) & 44.0 ($\bm{\alpha^{\star}}$=0.50) & 55.7 ($\bm{\alpha^{\star}}$=0.60) & 100.0 ($\bm{\alpha^{\star}}$=0.50) & 55.0 ($\bm{\alpha^{\star}}$=0.50) & 100.0 ($\bm{\alpha^{\star}}$=0.50) & 57.0 ($\bm{\alpha^{\star}}$=0.50) \\
CrossFire~\cite{dou2024adversarialattacksmultimodalmodels} & 79.4 ($\bm{\alpha^{\star}}$=0.99) & 54.3 ($\bm{\alpha^{\star}}$=0.99) & 53.7 ($\bm{\alpha^{\star}}$=1.00) & 82.0 ($\bm{\alpha^{\star}}$=0.99) & 54.0 ($\bm{\alpha^{\star}}$=0.99) & 78.0 ($\bm{\alpha^{\star}}$=0.99) & 51.0 ($\bm{\alpha^{\star}}$=0.99) \\
PTA~\cite{zhang2026improvinggeneralizabilityundetectabilitytargeted} & 100.0 ($\bm{\alpha^{\star}}$=0.70) & 43.1 ($\bm{\alpha^{\star}}$=0.60) & 53.7 ($\bm{\alpha^{\star}}$=0.50) & 100.0 ($\bm{\alpha^{\star}}$=0.50) & 54.0 ($\bm{\alpha^{\star}}$=0.50) & 100.0 ($\bm{\alpha^{\star}}$=0.50) & 55.0 ($\bm{\alpha^{\star}}$=0.70) \\
\textbf{\textsc{GapAttack}} & 100.0 ($\bm{\alpha^{\star}}$=0.50) & \textbf{57.4} ($\bm{\alpha^{\star}}$=0.50) & \textbf{64.9} ($\bm{\alpha^{\star}}$=0.50) & 100.0 ($\bm{\alpha^{\star}}$=0.50) & \textbf{62.0} ($\bm{\alpha^{\star}}$=0.60) & 100.0 ($\bm{\alpha^{\star}}$=0.50) & 58.0 ($\bm{\alpha^{\star}}$=0.60) \\
\textbf{\textsc{GapAttack}*} & 100.0 ($\bm{\alpha^{\star}}$=0.50) & 53.4 ($\bm{\alpha^{\star}}$=0.50) & 60.3 ($\bm{\alpha^{\star}}$=0.50) & 100.0 ($\bm{\alpha^{\star}}$=0.50) & 61.0 ($\bm{\alpha^{\star}}$=0.50) & 100.0 ($\bm{\alpha^{\star}}$=0.50) & \textbf{59.0} ($\bm{\alpha^{\star}}$=0.50) \\
\hline
\end{tabular}
\label{tab:openclip_results}
\end{center}
\vspace{-2em}
\end{table*}

\section{Experimental Evaluation}
\label{sec:experiments}

\subsection{Experimental Setup}

\textbf{Data and targets.} 
We evaluate targeted image-text adversarial alignment, where source images are perturbed to align their embeddings
with adversary-selected text targets. 
Source images are sampled from ImageNet-V2~\cite{recht2019imagenetclassifiersgeneralizeimagenet}, MS-COCO~\cite{10.1007/978-3-319-10602-1_48}, and STL-10~\cite{pmlr-v15-coates11a} using the 1{,}000 ImageNet class labels.
For MS-COCO and STL-10, we define the source class as the encoder's zero-shot top-$1$ prediction over these 1,000 classes.
The sampled splits and image counts are $350$ images for ImageNet-V2 and $100$ images each from MS-COCO and STL-10.
ImageBind attacks are run on a 100‑image ImageNet‑V2 cohort. For ImageBind, we additionally examine modality geometry using ImageNet-V2 and ESC-50~\cite{10.1145/2733373.2806390}, sampling two audio clips from each of ESC-50's 50 environmental-sound categories.

For each source image, we uniformly sample a target class from the remaining 999 ImageNet classes, excluding its source class.
We represent the selected target using the prompt ``A photo of a \{class\}''~\cite{radford2021clip}.
The same source--target pairs are used across all attacks.

\begin{table}[t]
\caption{Attack performance (\%) on Ablations of $\mathscr{G}$ .}
\begin{center}
\setlength{\tabcolsep}{5.3pt}
\begin{tabular}{|l|c|c|c|}
\hline
\rowcolor[gray]{0.88}
\textbf{Attack} & \textbf{\textit{Zero-shot}}
 & \textbf{\textit{COCO Cap. Gen.}} & \textbf{\textit{CC Cap. Gen.}} \\
\hline
Random Direction & 99.7  & 37.1  & 48.6 \\
\hline
No $\mathscr{G}$ Term & 100.0  & 44.0 & 55.7 \\
\hline
\textbf{\textsc{GapAttack}} & 100.0 & \textbf{57.4} & \textbf{64.9} \\
\hline
\end{tabular}
\label{tab:axis_ablation}
\end{center}
\vspace{-2em}
\end{table}

\textbf{Encoder Models.} 
We use OpenCLIP ViT-B/32 with the original OpenAI weights (\texttt{ViT-B-32-quickgelu}), producing 512-dimensional embeddings, and ImageBind-Huge~\cite{girdhar2023imagebind}, producing 1,024-dimensional embeddings across six modalities. Embeddings are unit-normalized. Attacks have white-box access to the encoder and perturb the input image under $\|\delta\|_{\infty}\le32/255$. 
The clean reference samples used to estimate the modality centroids are: for OpenCLIP, $9{,}000$ ImageNet-V2 images (disjoint from the attacked set) for the image centroid and the $1{,}000$ class prompts for the text centroid; for ImageBind, $100$ ImageNet-V2 images (vision), $100$ ESC-50 clips, two per category (audio), and $150$ prompts, one per distinct object and sound class (text).
For the axis used by each encoder, the standard deviation of clean-image projections, normalized by the corresponding centroid distance $\|g\|_2$, is $5.3\%$ for OpenCLIP and $11.7\%$ for ImageBind.

\textbf{Downstream Tasks.}
We evaluate the following tasks:
\begin{itemize}\itemsep1pt
  \item \emph{Zero-shot classification.} Images are classified by cosine similarity to the 1,000 embedded ImageNet class prompts (margin consumer).
  \item \emph{Captioning.} For OpenCLIP, we use ClipCap-COCO and ClipCap-CC~\cite{mokady2021clipcapclipprefiximage}, trained on MS-COCO and Conceptual Captions, respectively. Both use a two-layer MLP to map a 512-dimensional image embedding to a ten-token prefix for a GPT-2 decoder.
For ImageBind we use IBCap, a  ClipCap-style decoder we train: a two-layer MLP ($1024\!\to\!3840\!\to\!7680$, Tanh) maps the unit-normalized $1{,}024$-dimensional ImageBind embedding to a length-$10$ prefix of GPT-2 token embeddings, which GPT-2 decodes into a caption. It is trained on MS-COCO \texttt{train2017} ($118{,}287$ images, $591{,}753$ captions), and both attack arms are judged by this same decoder.
  \item \emph{Retrieval}. An adversarial ImageBind embedding queries a fixed gallery of 1,000 real images, with one image per ImageNet class. 
  \item \emph{Image Generation}. We use BindDiffusion~\cite{binddiffusion2023} to condition an unCLIP-style diffusion pipeline~\cite{ramesh2022hierarchicaltextconditionalimagegeneration} on ImageBind embeddings.
\end{itemize}

\textbf{Metrics.} 
Classification succeeds when the predicted class equals the chosen target; retrieval succeeds when the top-ranked gallery image belongs to that class.
For captioning, generated captions are embedded with the OpenCLIP ViT-B/32 text encoder (OpenAI weights, \texttt{ViT-B-32-quickgelu}) and compared with the 1{,}000 embedded class prompts; this is the same text encoder throughout, and for ImageBind it is independent of the attacked encoder.
An attack succeeds when the target class ranks among the top ten.
The corresponding clean-input rates are $2.0\%$ for ClipCap-COCO and $1.7\%$ for ClipCap-CC.
Image-generation success is determined by the same CLIP ViT-B/32 zero-shot classifier used for the other consumers: the generated image is classified against the $1{,}000$ class prompts and the attack succeeds when the target class is within the top ten; we generate one image per embedding (a single deterministic sample, the latent noise seeded per source index so both arms share identical noise).

Success at an alignment threshold is the fraction of the full cohort that succeeds, with inputs not reaching the threshold counted as failures. For captioning, a threshold is evaluated only when at least 20 inputs reach it; thresholds reached by fewer are reported as N/A. Retrieval is scored over all reached inputs and therefore reports a value at every threshold.
Downstream outputs are used only for evaluation, not for attack construction or calibration of $\lambda$.

\textbf{Implementation Details.}
Experiments run on an NVIDIA GeForce RTX~3070 with 8\,GB memory, using PyTorch~2.6.0 with CUDA~12.4. PGD~\cite{madry2018towards} starts from $\delta_0=0$ and uses sign-gradient updates with a step size decayed from $\epsilon/4$ to $\epsilon/100$ over at most $7{,}500$ iterations, subject to $\|\delta\|_{\infty}\le\epsilon=32/255$.
We calibrate $\lambda$ using the encoder-only procedure in Section~\ref{sec:attack}, obtaining $\lambda=0.5$ for OpenCLIP and $\lambda=0.4$ for ImageBind.
The calibrated weight is reused across attacks on the same encoder.

\subsection{Evaluation on OpenCLIP}
\subsubsection{Baseline Attacks}
We compare \textsc{GapAttack} with Illusions~\cite{zhang2024illusions}, CrossFire~\cite{dou2024adversarialattacksmultimodalmodels}, and PTA~\cite{zhang2026improvinggeneralizabilityundetectabilitytargeted}. These methods construct targeted attacks on shared multi-modal embeddings for transfer to downstream consumers, making them relevant baselines for our setting. All methods are evaluated using the same source--target pairs and perturbation budget.
To assess whether proxy-based alignment provides additional gains, we also evaluate \textsc{GapAttack}*, a variant that aligns with multiple target proxies following PTA.

\begin{table*}[!t]
\caption{\textbf{\textsc{GapAttack}'s Joint Alignment versus Alignment-only:} Downstream success (\%) at each alignment level $\alpha$ on ImageBind.}
\vspace{-1.5em}
\begin{center}
\small
\setlength{\tabcolsep}{6.6pt}
\begin{tabular}{l||c|c|c|c|c||c|c|c|c|c}
\hline
\rowcolor[gray]{0.75}
 & \multicolumn{5}{c||}{Caption Generation (ImageBind Decoder)} & \multicolumn{5}{c}{Image Generation (BindDiffusion)} \\
\hhline{-||-|-|-|-|-||-|-|-|-|-}
Alignment Type & $\alpha{=}0.3$ & $\alpha{=}0.4$ & $\alpha{=}0.5$ & $\alpha{=}0.6$ & $\alpha{=}0.7$ & $\alpha{=}0.3$ & $\alpha{=}0.4$ & $\alpha{=}0.5$ & $\alpha{=}0.6$ & $\alpha{=}0.7$ \\
\hhline{=#=|=|=|=|=#=|=|=|=|=}
Alignment-only~\cite{zhang2024illusions, dou2024adversarialattacksmultimodalmodels, zhang2026improvinggeneralizabilityundetectabilitytargeted} & 22.0 & 38.0 & 44.0 & 48.0 & 45.0 & 24.0 & 57.0 & 69.0 & 80.0 & 81.0 \\
\textbf{Joint Alignment (Ours)} & \textbf{24.0} & \textbf{47.0} & \textbf{49.0} & \textbf{51.0} & \textbf{56.0} & \textbf{30.0} & \textbf{60.0} & \textbf{82.0} & \textbf{84.0} & \textbf{90.0} \\
\hline
\end{tabular}
\label{tab:alpha_sweep}
\end{center}
\vspace{-2em}
\end{table*}

\subsubsection{{Attack Performance}} 
We evaluate targeted attack success at prescribed alignment thresholds between $0.1$ and $1.0$. Table~\ref{tab:openclip_results} reports results at $\alpha^\star$, the evaluated checkpoint with the highest observed targeted attack success rate for each method, dataset, and downstream consumer. This checkpoint is selected post hoc for comparison and does not guide attack construction or calibration of $\lambda$. Reported alignment is the cosine similarity between the adversarial input embedding and each attack's target: the text embedding for Illusions, PTA, and GapAttack, and the generated image embedding for CrossFire. CrossFire's alignment is therefore measured on the image side of the modality gap and is not on the same scale as the text‑target methods.

\textsc{GapAttack} outperforms all three external baselines in the four evaluated captioning settings, with gains of $1.0$--$9.2$ percentage points over the strongest competing baseline in each setting. It also achieves $100\%$ targeted classification success on all three datasets, matching Illusions and PTA. The proxy variant improves STL-10 captioning by one percentage point over \textsc{GapAttack}, but performs worse in the other three captioning settings. These results show that \textsc{GapAttack} achieves captioning gains across the evaluated settings using a single text target.

\subsubsection{Effect of Embedding Placement along Modality Axis} 
\begin{figure}[!t]
    \centering
    \includegraphics[width=\columnwidth]{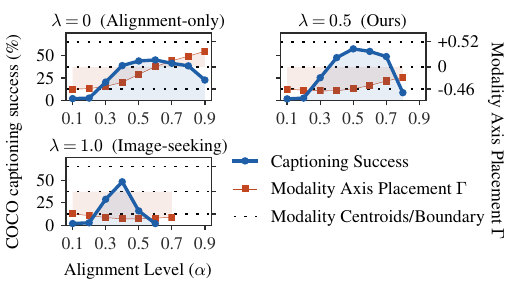}
    \vspace{-2em}
    \caption{\textbf{Effect of joint alignment versus alignment-only attacks, where $\lambda$ controls the adversarial embedding’s placement along the modality axis.}}
    \label{fig:lambda_curves}
    \vspace{-1.5em}
\end{figure}

Fig.~\ref{fig:lambda_curves} compares $\lambda=0$, $0.5$, and $1.0$ to examine how the placement weight affects embedding geometry and captioning attack success. For an adversarial embedding $z=\theta^{(m)}(x_m+\delta)$, we report the centroid similarity contrast $\Gamma(z)=\langle z,\hat\mu_{\mathrm{txt}}\rangle-\langle z,\hat\mu_{\mathrm{img}}\rangle$, where $\mu_i$ is the clean reference centroid of modality $i$ and $\hat\mu_i=\mu_i/\|\mu_i\|_2$. This diagnostic measures relative similarity to the two centroids; the attack objective uses the modality axis $\hat g$ defined in Eq.~\eqref{eq:axis}. The clean-image and text-target reference values are approximately $\Gamma=-0.46$ and $\Gamma=+0.52$, respectively. As target alignment increases, the alignment-only attack ($\lambda=0$) moves toward the text-side reference in this diagnostic. Compared with alignment-only optimization, the calibrated weight $\lambda=0.5$ favors more image-side placement and achieves the highest peak captioning success among the three settings, at $57.4\%$. Increasing $\lambda$ to $1.0$ yields a lower peak, indicating that a stronger placement bias does not necessarily improve attack success. These results support balancing target alignment with embedding placement.

\subsection{Ablation Studies}

Tables~\ref{tab:axis_ablation} and~\ref{tab:region} use the OpenCLIP ViT-B/32 encoder on ImageNet-V2 and follow the convention of Table~\ref{tab:openclip_results}, each cell reports the best alignment $\alpha^*$ over the 350-image cohort.
\subsubsection{Modality-axis term} We compare \textsc{GapAttack} with two variants: replacing the modality axis with a random direction, and removing the $\mathscr{G}$ term to recover alignment-only optimization. As shown in Table~\ref{tab:axis_ablation}, \textsc{GapAttack} improves captioning success over the alignment-only variant by $13.4$ percentage points on ClipCap-COCO and $9.2$ points on ClipCap-CC. The random-direction variant performs worse than alignment-only optimization on both captioners, while classification success remains near saturation across all three variants. These results support the contribution of the modality-axis term to captioning attack success in the evaluated setting.

\subsubsection{Consumer training modality} To examine the relationship between embedding placement and the modality used to train the consumer, we evaluate TextCap, a ClipCap-style captioner trained on CLIP text embeddings, motivated by text-only captioning~\cite{https://doi.org/10.48448/n7sq-p557}. We compare image-side placement under a positive $\lambda$ with text-side placement under $\lambda=-0.5$, using the centroid similarity contrast $\Gamma$ defined above to characterize their positions. The text-side condition yields $\Gamma\approx+0.49$, close to the text-target reference of approximately $+0.52$. At the reported checkpoints, image-side placement achieves higher attack success on all three captioners (Table~\ref{tab:region}), including the text-trained TextCap, for which success is $72.0\%$ versus $57.4\%$. Thus, proximity to the text-side reference in this diagnostic does not guarantee higher attack success on a consumer trained on text embeddings.

\begin{table}[!t]
\caption{Captioning attack success (\%) under image-side and text-side placement; $\Gamma$ denotes the centroid similarity contrast.}
\vspace{-1em}
\begin{center}\setlength{\tabcolsep}{0.7pt}\footnotesize
\begin{tabular}{|l|c|c|c|c|}
\hline
\rowcolor[gray]{0.88}
\textbf{Placement} & \textbf{$\Gamma$} & \textbf{ClipCap-COCO}
 & \textbf{ClipCap-CC} & \textbf{TextCap}\\
\rowcolor[gray]{0.88}
   of AE &   & \emph{(image-trained)} & \emph{(image-trained)} & \emph{(text-trained)}\\
\hline
Image side ($\lambda>0$) & $-0.46$ & \textbf{57.4} & \textbf{64.9} & \textbf{72.0}\\
\hline
Text side ($\lambda<0$)  & $+0.49$ & 36.0 & 38.6 & 57.4\\
\hline
\end{tabular}\end{center}\label{tab:region}
\vspace{-2.5em}
\end{table}

\subsection{Evaluation on ImageBind}
On ImageBind, we compare \textsc{GapAttack} with its alignment-only variant ($\lambda=0$) to assess the contribution of modality-axis control on a second encoder. We construct the modality axis from the clean image centroid $\mu_{\mathrm{img}}$ to the aggregate reference centroid $\mu_{\mathrm{ref}}$ of the other evaluated modalities, using $\hat g=(\mu_{\mathrm{ref}}-\mu_{\mathrm{img}})/\|\mu_{\mathrm{ref}}-\mu_{\mathrm{img}}\|_2$ ($\mu_{\mathrm{ref}}$ is the unweighted mean of the text and audio centroids). This extends the pairwise construction in Eq.~\eqref{eq:axis} by replacing the target-modality centroid with an aggregate reference centroid. Encoder-based calibration following Section~\ref{sec:attack} yields $\lambda=0.4$. Both variants use the same source--target pairs, perturbation budget, and iteration budget, without additional defensive preprocessing.

Table~\ref{tab:alpha_sweep} reports targeted captioning and image-generation success at five prescribed alignment thresholds from $\alpha=0.3$ to $0.7$. \textsc{GapAttack} improves both tasks at every reported threshold. The largest gains are $11.0$ percentage points for captioning at $\alpha=0.7$ and $13.0$ points for image generation at $\alpha=0.5$. In the evaluated zero-shot classification setting, both variants reach $100\%$ targeted success at $\alpha=0.5$.

\begin{table}[b]
\vspace{-2em}
\caption{Retrieval@1 (\%) on ImageBind across alignment levels $\alpha$.}
\vspace{-1.5em}
\begin{center}
\small
\setlength{\tabcolsep}{1.4pt}
\begin{tabular}{l||c|c|c|c|c|c|c|c|c}
\hline
\rowcolor[gray]{0.75}
$\alpha$ & 0.1 & 0.2 & 0.3 & 0.4 & 0.5 & 0.6 & 0.7 & 0.8 & 0.9 \\
\hhline{=#=|=|=|=|=|=|=|=|=}
Alignment-only ($\lambda{=}0$) & 1.0 & 2.0 & 14.0 & \textbf{43.0} & 44.0 & 50.0 & \textbf{62.0} & \textbf{63.0} & \textbf{57.0} \\
Joint (Ours)($\lambda{=}0.4$) & 1.0 & 1.0 & \textbf{17.0} & 31.0 & \textbf{45.0} & \textbf{51.0} & 53.0 & 52.0 & 2.0 \\
\hline
\end{tabular}
\label{tab:retrieval_alpha}
\end{center}
\vspace{-2em}
\end{table}
\paragraph{Retrieval} Table~\ref{tab:retrieval_alpha} shows mixed results: \textsc{GapAttack} achieves higher success than alignment-only optimization at three of the nine alignment thresholds, ties at one, and performs worse at five. Retrieval depends on the target gallery image's rank relative to other gallery images, which is not determined solely by alignment with the text target. Together with the consistent captioning and image-generation gains reported above, these results show that the benefit of modality-axis control varies across downstream consumers. At high alignment thresholds, full-cohort retrieval success also reflects differences in threshold attainment within the iteration budget.

\begin{figure}[t]
    \centering
    \includegraphics[width=\columnwidth]{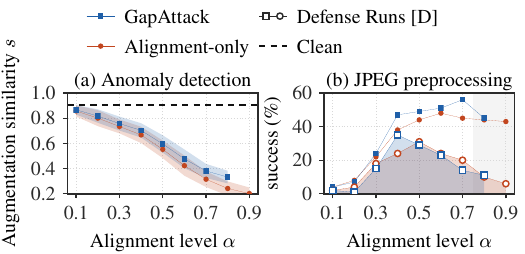}
    \vspace{-2em}

    \caption{\textbf{Embedding stability under augmentations (a) and captioning attack success with and without JPEG-preprocessing (b) on ImageBind. Bands in (a) show one standard deviation.} 
    }
    \label{fig:def_curves}
    \vspace{-1em}
\end{figure}

\subsection{Evaluation Against Defenses}
\label{sec:jpeg_defense}
We examine how input transformations used in defenses affect adversarial embedding stability and downstream attack success on ImageBind. Both \textsc{GapAttack} and alignment-only optimization are evaluated without adapting their objectives to these transformations.

\paragraph{Augmentation-based stability}
Augmentation-based anomaly detectors use inconsistencies between an input's embedding and those of its transformed versions as a detection signal~\cite{zhang2024illusions}. We quantify embedding stability by the augmentation similarity $s(x)=\mathbb{E}_a\!\left[\cos\!\big(\theta(x),\theta(a(x))\big)\right]$ for a clean input $x$ and apply the same definition to its adversarial counterpart $x_\delta=x+\delta$, yielding $s(x_\delta)$. We use the $100$ image cohort and six augmentations, including JPEG compression (quality $75$), Gaussian blur (kernel $5$, $\sigma{=}1.5$), random affine ($10^{\circ}$ rotation, $0.05$ translation, $0.95$ scaling, $5^{\circ}$ shear), color jitter (brightness $0.80$, contrast $0.80$, saturation $1.40$, hue $0.15$), horizontal flip (parameter-free), and perspective warp (distortion $0.25$).
For the adversarial curves, means and standard deviations are computed over inputs reaching each alignment threshold.

As shown in Fig.~\ref{fig:def_curves}(a), \textsc{GapAttack} achieves higher mean augmentation similarity than alignment-only optimization at every jointly evaluated threshold. At $\alpha{=}0.7$, the mean scores are $0.38$ and $0.32$, respectively, compared with $0.91$ for clean inputs (Table~\ref{tab:aug_sim}). These results show a relative improvement in embedding stability, although both attacks remain substantially less stable than clean inputs under the tested transformations.

\paragraph{JPEG preprocessing}
To assess the downstream effect of input transformation, we evaluate targeted captioning success after JPEG compression~\cite{Liu_2019_CVPR} at quality $75$. 
Compression is applied to adversarial checkpoint inputs before encoding and $\alpha$ denotes the pre-compression alignment threshold. We focus on the five alignment levels from $0.3$ to $0.7$ used in Table~\ref{tab:alpha_sweep}, which both attacks reach on the full cohort and which cover their observed success peaks.
JPEG preprocessing substantially reduces captioning success for both attacks, while \textsc{GapAttack} retains a higher observed peak than alignment-only optimization ($35.0\%$ versus $31.0\%$; Fig.~\ref{fig:def_curves}(b)). Across the five thresholds, however, mean success rates are numerically similar ($23.2\%$ and $23.4\%$, respectively), indicating that the retained advantage depends on the checkpoint under this JPEG configuration.

\begin{table}[t]
\caption{Augmentation similarity $s$ at $\alpha{=}0.7$ on ImageBind: clean input $x$ versus the perturbed inputs of the two attacks.}
\vspace{-1.5em}
\begin{center}
\small
\setlength{\tabcolsep}{4.6pt}
\begin{tabular}{l||c|c|c}
\hline
\rowcolor[gray]{0.85}
Augmentation & $x$ & Joint $x_\delta$ (Ours) & Alignment-only $x_\delta$  \\
\hhline{=#=|=|=}
JPEG compression   & 0.93 & \textbf{0.50} & 0.49 \\
Gaussian blur      & 0.91 & \textbf{0.30} & 0.19 \\
Random affine      & 0.91 & \textbf{0.32} & 0.23 \\
Color jitter       & 0.87 & \textbf{0.41} & 0.38 \\
Horizontal flip    & 0.97 & \textbf{0.44} & 0.40 \\
Perspective warp   & 0.84 & \textbf{0.31} & 0.21 \\
\hline
Average            & 0.91 & \textbf{0.38} & 0.32 \\
\hline
\end{tabular}
\label{tab:aug_sim}
\end{center}
\vspace{-2.5em}
\end{table}

\subsection{Practical Considerations and Limitations}
\label{sec:practical}

\textsc{GapAttack} constructs attacks and calibrates $\lambda$ using white-box access to the fixed encoder and available clean reference samples, without querying downstream consumers. For $d$-dimensional embeddings, the modality-axis term adds $O(d)$ arithmetic per iteration using the embedding already computed for target alignment, so it requires no additional encoder forward pass. 
Centroid estimation and calibration incur a one-time setup cost, after which the axis and weight are reused.
Our evaluation covers digital image perturbations on two encoders under an $\ell_\infty$ budget of $32/255$, with success at high alignment thresholds additionally limited by whether optimization reaches the prescribed threshold within budget. These results therefore require interpretation alongside threshold-attainment rates, while their sensitivity to preprocessing is examined through JPEG compression in Section~\ref{sec:jpeg_defense}.

\paragraph{Ethical considerations} Our evaluation is restricted to public benchmarks and open encoders, with the goal of characterizing vulnerabilities in shared multi-modal representations and informing robustness assessment.

\section{Conclusion}
\label{sec:conclusion}

We presented \textsc{GapAttack}, which jointly optimizes target alignment and embedding placement along a centroid-defined modality axis without querying downstream consumers. Experiments on OpenCLIP and ImageBind show improved targeted captioning and image-generation success, including gains over alignment-only optimization at prescribed alignment thresholds. These findings highlight embedding placement as a factor beyond target alignment in downstream vulnerability, with benefits that vary across consumers and preprocessing conditions.

\section*{Acknowledgment}
This work was partially supported by the National Science Foundation (NSF) under Grant No. 2531855.

\bibliographystyle{IEEEtran}
\bibliography{ref}

@String(CVPR= {IEEE Conf. Comput. Vis. Pattern Recog.})

@String(ICCV= {Int. Conf. Comput. Vis.})

@String(ECCV= {Eur. Conf. Comput. Vis.})

@String(ICLR = {Int. Conf. Learn. Represent.})

@String(CVPR  = {CVPR})

@String(ICCV  = {ICCV})

@String(ECCV  = {ECCV})

@String(ICLR  = {ICLR})

@inproceedings{zhou2025runtime,
  title={Runtime stealthy perception attacks against dnn-based adaptive cruise control systems},
  author={Zhou, Xugui and Chen, Anqi and Kouzel, Maxfield and Ren, Haotian and McCarty, Morgan and Nita-Rotaru, Cristina and Alemzadeh, Homa},
  booktitle={Proceedings of the 20th ACM Asia Conference on Computer and Communications Security},
  pages={1065--1082},
  year={2025}
}

@inproceedings{chen2025revisiting,
  title={Revisiting adversarial perception attacks and defense methods on autonomous driving systems},
  author={Chen, Cheng and Wang, Yuhong and Munir, Nafis S and Zhou, Xiangwei and Zhou, Xugui},
  booktitle={2025 55th Annual IEEE/IFIP International Conference on Dependable Systems and Networks Workshops (DSN-W)},
  pages={242--249},
  year={2025},
  organization={IEEE}
}

@inproceedings{zhou2022robustness,
  title={Robustness testing of data and knowledge driven anomaly detection in cyber-physical systems},
  author={Zhou, Xugui and Kouzel, Maxfield and Alemzadeh, Homa},
  booktitle={2022 52nd Annual IEEE/IFIP International Conference on Dependable Systems and Networks Workshops (DSN-W)},
  pages={44--51},
  year={2022},
  organization={IEEE}
}

@inproceedings{radford2021clip,
  title     = {Learning Transferable Visual Models From Natural Language Supervision},
  author    = {Radford, Alec and Kim, Jong Wook and Hallacy, Chris and
               Ramesh, Aditya and Goh, Gabriel and Agarwal, Sandhini and
               Sastry, Girish and Askell, Amanda and Mishkin, Pamela and
               Clark, Jack and Krueger, Gretchen and Sutskever, Ilya},
  booktitle = {International Conference on Machine Learning (ICML)},
  year      = {2021}
}

@inproceedings{girdhar2023imagebind,
  title     = {ImageBind: One Embedding Space To Bind Them All},
  author    = {Girdhar, Rohit and El-Nouby, Alaaeldin and Liu, Zhuang and
               Singh, Mannat and Alwala, Kalyan Vasudev and
               Joulin, Armand and Misra, Ishan},
  booktitle = {IEEE/CVF Conference on Computer Vision and Pattern Recognition (CVPR)},
  year      = {2023}
}

@inproceedings{zhang2024illusions,
  title     = {Adversarial Illusions in Multi-Modal Embeddings},
  author    = {Zhang, Tingwei and Jha, Rishi and Bagdasaryan, Eugene and
               Shmatikov, Vitaly},
  booktitle = {USENIX Security Symposium},
  year      = {2024},
  note      = {Distinguished Paper Award}
}

@inproceedings{zhang2026hubness,
  title     = {Adversarial Hubness in Multi-Modal Retrieval},
  author    = {Zhang, Tingwei and Suya, Fnu and Jha, Rishi and
               Zhang, Collin and Shmatikov, Vitaly},
  booktitle = {IEEE Symposium on Security and Privacy (S\&P)},
  year      = {2026}
}

@inproceedings{zhang2025selfinterpreting,
  title     = {Self-interpreting Adversarial Images},
  author    = {Zhang, Tingwei and Zhang, Collin and Morris, John X. and
               Bagdasarian, Eugene and Shmatikov, Vitaly},
  booktitle = {USENIX Security Symposium},
  year      = {2025}
}

@inproceedings{shayegani2024plugandpray,
  title     = {Jailbreak in Pieces: Compositional Adversarial Attacks on
               Multi-Modal Language Models},
  author    = {Shayegani, Erfan and Dong, Yue and Abu-Ghazaleh, Nael},
  booktitle = {International Conference on Learning Representations (ICLR)},
  year      = {2024},
  note      = {Earlier version circulated as
               ``Plug and Pray: Exploiting off-the-shelf components of
               Multi-Modal Models'', arXiv:2307.14539}
}

@inproceedings{schlarmann2024robustclip,
  title     = {Robust {CLIP}: Unsupervised Adversarial Fine-Tuning of Vision
               Embeddings for Robust Large Vision-Language Models},
  author    = {Schlarmann, Christian and Singh, Naman Deep and
               Croce, Francesco and Hein, Matthias},
  booktitle = {International Conference on Machine Learning (ICML)},
  year      = {2024}
}

@misc{liu2025bridgingtaskgapmultitask,
      title={Bridging the Task Gap: Multi-Task Adversarial Transferability in CLIP and Its Derivatives}, 
      author={Kuanrong Liu and Siyuan Liang and Cheng Qian and Ming Zhang and Xiaochun Cao},
      year={2025},
      eprint={2509.23917},
      archivePrefix={arXiv},
      primaryClass={cs.CV},
      url={https://arxiv.org/abs/2509.23917}, 
}

@misc{silva2025attackingattentionfoundationmodels,
      title={Attacking Attention of Foundation Models Disrupts Downstream Tasks}, 
      author={Hondamunige Prasanna Silva and Federico Becattini and Lorenzo Seidenari},
      year={2025},
      eprint={2506.05394},
      archivePrefix={arXiv},
      primaryClass={cs.CR},
      url={https://arxiv.org/abs/2506.05394}, 
}

@misc{zhang2026improvinggeneralizabilityundetectabilitytargeted,
      title={Improving Generalizability and Undetectability for Targeted Adversarial Attacks on Multimodal Pre-trained Models}, 
      author={Zhifang Zhang and Jiahan Zhang and Shengjie Zhou and Qi Wei and Shuo He and Feng Liu and Lei Feng},
      year={2026},
      eprint={2509.19994},
      archivePrefix={arXiv},
      primaryClass={cs.CV},
      url={https://arxiv.org/abs/2509.19994}, 
}

@misc{akbarian2026breakingillusionconsensusbasedgenerative,
      title={Breaking the Illusion: Consensus-Based Generative Mitigation of Adversarial Illusions in Multi-Modal Embeddings}, 
      author={Fatemeh Akbarian and Anahita Baninajjar and Yingyi Zhang and Ananth Balashankar and Amir Aminifar},
      year={2026},
      eprint={2511.21893},
      archivePrefix={arXiv},
      primaryClass={cs.LG},
      url={https://arxiv.org/abs/2511.21893}, 
}

@inproceedings{
madry2018towards,
title={Towards Deep Learning Models Resistant to Adversarial Attacks},
author={Aleksander Madry and Aleksandar Makelov and Ludwig Schmidt and Dimitris Tsipras and Adrian Vladu},
booktitle={International Conference on Learning Representations},
year={2018},
url={https://openreview.net/forum?id=rJzIBfZAb},
}

@misc{zhao2023evaluatingadversarialrobustnesslarge,
      title={On Evaluating Adversarial Robustness of Large Vision-Language Models}, 
      author={Yunqing Zhao and Tianyu Pang and Chao Du and Xiao Yang and Chongxuan Li and Ngai-Man Cheung and Min Lin},
      year={2023},
      eprint={2305.16934},
      archivePrefix={arXiv},
      primaryClass={cs.CV},
      url={https://arxiv.org/abs/2305.16934}, 
}

@misc{liang2022mindgapunderstandingmodality,
      title={Mind the Gap: Understanding the Modality Gap in Multi-modal Contrastive Representation Learning}, 
      author={Weixin Liang and Yuhui Zhang and Yongchan Kwon and Serena Yeung and James Zou},
      year={2022},
      eprint={2203.02053},
      archivePrefix={arXiv},
      primaryClass={cs.CL},
      url={https://arxiv.org/abs/2203.02053}, 
}

@misc{mokady2021clipcapclipprefiximage,
      title={ClipCap: CLIP Prefix for Image Captioning}, 
      author={Ron Mokady and Amir Hertz and Amit H. Bermano},
      year={2021},
      eprint={2111.09734},
      archivePrefix={arXiv},
      primaryClass={cs.CV},
      url={https://arxiv.org/abs/2111.09734}, 
}

@misc{daw2024hidinginplainsighthipsattackclip,
      title={Hiding-in-Plain-Sight (HiPS) Attack on CLIP for Targetted Object Removal from Images}, 
      author={Arka Daw and Megan Hong-Thanh Chung and Maria Mahbub and Amir Sadovnik},
      year={2024},
      eprint={2410.13010},
      archivePrefix={arXiv},
      primaryClass={cs.LG},
      url={https://arxiv.org/abs/2410.13010}, 
}

@misc{chowers2026modalitygapbugfeature,
      title={Is the Modality Gap a Bug or a Feature? A Robustness Perspective}, 
      author={Rhea Chowers and Oshri Naparstek and Udi Barzelay and Yair Weiss},
      year={2026},
      eprint={2603.29080},
      archivePrefix={arXiv},
      primaryClass={cs.CV},
      url={https://arxiv.org/abs/2603.29080}, 
}

@inproceedings{zhang2025clipure,
  title={{CLIPure}: Purification in Latent Space via {CLIP} for Adversarially
         Robust Zero-Shot Classification},
  author={Zhang, Mingkun and Bi, Keping and Chen, Wei and Guo, Jiafeng and
          Cheng, Xueqi},
  booktitle={ICLR}, year={2025}, note={arXiv:2502.18176}}

@inproceedings{krakover2026a4d,
  title={A Classifier-Agnostic Zero-Shot Adversarial Attack Detection via {CLIP}},
  author={Krakover, Hodaya and Levi, Meir Yossef and Gofer, Eyal and Gilboa, Guy},
  booktitle={ECCV}, series={LNCS}, volume={17055}, pages={415--431},
  year={2026}, doi={10.1007/978-3-032-37261-1_24}}

@inproceedings{zhu2025cola,
  title={Enhancing {CLIP} Robustness via Cross-Modality Alignment},
  author={Zhu, Xingyu and Zhu, Beier and Wang, Shuo and Zhao, Kesen and
          Zhang, Hanwang},
  booktitle={NeurIPS}, year={2025}, note={arXiv:2510.24038}}

@inproceedings{https://doi.org/10.48448/n7sq-p557,
  title={Text-only training for image captioning using noise-injected clip},
  author={Nukrai, David and Mokady, Ron and Globerson, Amir},
  booktitle={Findings of the Association for Computational Linguistics: EMNLP 2022},
  pages={4055--4063},
  year={2022}
}

@misc{wang2022understandingcontrastiverepresentationlearning,
      title={Understanding Contrastive Representation Learning through Alignment and Uniformity on the Hypersphere}, 
      author={Tongzhou Wang and Phillip Isola},
      year={2020},
      eprint={2005.10242},
      archivePrefix={arXiv},
      primaryClass={cs.LG},
      url={https://arxiv.org/abs/2005.10242}, 
}

@misc{you2022learningvisualrepresentationmodalityshared,
      title={Learning Visual Representation from Modality-Shared Contrastive Language-Image Pre-training}, 
      author={Haoxuan You and Luowei Zhou and Bin Xiao and Noel Codella and Yu Cheng and Ruochen Xu and Shih-Fu Chang and Lu Yuan},
      year={2022},
      eprint={2207.12661},
      archivePrefix={arXiv},
      primaryClass={cs.CV},
      url={https://arxiv.org/abs/2207.12661}, 
}

@misc{recht2019imagenetclassifiersgeneralizeimagenet,
      title={Do ImageNet Classifiers Generalize to ImageNet?}, 
      author={Benjamin Recht and Rebecca Roelofs and Ludwig Schmidt and Vaishaal Shankar},
      year={2019},
      eprint={1902.10811},
      archivePrefix={arXiv},
      primaryClass={cs.CV},
      url={https://arxiv.org/abs/1902.10811}, 
}

@inproceedings{
grassucci2025closing,
title={Closing The Modality Gap Enables Novel Multimodal Learning Applications},
author={Eleonora Grassucci and Giordano Cicchetti and Danilo Comminiello},
booktitle={Second Workshop on Representational Alignment at ICLR 2025},
year={2025},
url={https://openreview.net/forum?id=P3Oba8Z3B6}
}

@misc{fonseca2025textonlytrainingimagecaptioning,
      title={Text-Only Training for Image Captioning with Retrieval Augmentation and Modality Gap Correction}, 
      author={Rui Fonseca and Bruno Martins and Gil Rocha},
      year={2025},
      eprint={2512.04309},
      archivePrefix={arXiv},
      primaryClass={cs.CV},
      url={https://arxiv.org/abs/2512.04309}, 
}

@misc{dou2024adversarialattacksmultimodalmodels,
      title={Adversarial Attacks to Multi-Modal Models}, 
      author={Zhihao Dou and Xin Hu and Haibo Yang and Zhuqing Liu and Minghong Fang},
      year={2024},
      eprint={2409.06793},
      archivePrefix={arXiv},
      primaryClass={cs.CR},
      url={https://arxiv.org/abs/2409.06793}, 
}

@misc{inkawhich2023adversarialattacksfoundationalvision,
      title={Adversarial Attacks on Foundational Vision Models}, 
      author={Nathan Inkawhich and Gwendolyn McDonald and Ryan Luley},
      year={2023},
      eprint={2308.14597},
      archivePrefix={arXiv},
      primaryClass={cs.CV},
      url={https://arxiv.org/abs/2308.14597}, 
}

@InProceedings{10.1007/978-3-319-10602-1_48,
author="Lin, Tsung-Yi
and Maire, Michael
and Belongie, Serge
and Hays, James
and Perona, Pietro
and Ramanan, Deva
and Doll{\'a}r, Piotr
and Zitnick, C. Lawrence",
editor="Fleet, David
and Pajdla, Tomas
and Schiele, Bernt
and Tuytelaars, Tinne",
title="Microsoft COCO: Common Objects in Context",
booktitle="Computer Vision -- ECCV 2014",
year="2014",
publisher="Springer International Publishing",
address="Cham",
pages="740--755",
isbn="978-3-319-10602-1"
}

@InProceedings{pmlr-v15-coates11a,
  title = 	 {An Analysis of Single-Layer Networks in Unsupervised Feature Learning},
  author = 	 {Coates, Adam and Ng, Andrew and Lee, Honglak},
  booktitle = 	 {Proceedings of the Fourteenth International Conference on Artificial Intelligence and Statistics},
  pages = 	 {215--223},
  year = 	 {2011},
  editor = 	 {Gordon, Geoffrey and Dunson, David and Dudík, Miroslav},
  volume = 	 {15},
  series = 	 {Proceedings of Machine Learning Research},
  address = 	 {Fort Lauderdale, FL, USA},
  month = 	 {11--13 Apr},
  publisher =    {PMLR},
  url = 	 {https://proceedings.mlr.press/v15/coates11a.html}
}

@inproceedings{10.1145/2733373.2806390,
author = {Piczak, Karol J.},
title = {ESC: Dataset for Environmental Sound Classification},
year = {2015},
isbn = {9781450334594},
publisher = {Association for Computing Machinery},
address = {New York, NY, USA},
url = {https://doi.org/10.1145/2733373.2806390},
doi = {10.1145/2733373.2806390},
booktitle = {Proceedings of the 23rd ACM International Conference on Multimedia},
pages = {1015–1018},
numpages = {4},
location = {Brisbane, Australia},
series = {MM '15}
}

@article{zhou2022coop,
  title   = {Learning to Prompt for Vision-Language Models},
  author  = {Zhou, Kaiyang and Yang, Jingkang and Loy, Chen Change and Liu, Ziwei},
  journal = {International Journal of Computer Vision},
  volume  = {130},
  number  = {9},
  pages   = {2337--2348},
  year    = {2022}
}

@misc{ramesh2022hierarchicaltextconditionalimagegeneration,
      title={Hierarchical Text-Conditional Image Generation with CLIP Latents}, 
      author={Aditya Ramesh and Prafulla Dhariwal and Alex Nichol and Casey Chu and Mark Chen},
      year={2022},
      eprint={2204.06125},
      archivePrefix={arXiv},
      primaryClass={cs.CV},
      url={https://arxiv.org/abs/2204.06125}, 
}

@inproceedings{liu2023llava,
  title     = {Visual Instruction Tuning},
  author    = {Liu, Haotian and Li, Chunyuan and Wu, Qingyang and Lee, Yong Jae},
  booktitle = {Advances in Neural Information Processing Systems (NeurIPS)},
  volume    = {36},
  year      = {2023}
}

@misc{binddiffusion2023,
  title        = {{BindDiffusion}: One Diffusion Model to Bind Them All},
  author       = {Lin, Zhijie and Liu, Lijuan and Wang, Yangzihao and Xu, Xiangyu},
  year         = {2023},
  howpublished = {\url{https://github.com/sail-sg/BindDiffusion/tree/main}},
  note         = {GitHub repository}
}

@misc{fahim2024itsmodalitygapcharacterizing,
      title={It's Not a Modality Gap: Characterizing and Addressing the Contrastive Gap}, 
      author={Abrar Fahim and Alex Murphy and Alona Fyshe},
      year={2024},
      eprint={2405.18570},
      archivePrefix={arXiv},
      primaryClass={cs.CV},
      url={https://arxiv.org/abs/2405.18570}, 
}

@inproceedings{eslami2025mitigate,
  title     = {Mitigate the Gap: Improving Cross-Modal Alignment in {CLIP}},
  author    = {Eslami, Sedigheh and de Melo, Gerard},
  booktitle = {International Conference on Learning Representations (ICLR)},
  year      = {2025}
}

@InProceedings{Huang_2025_ICCV,
    author    = {Huang, Linlan and Cao, Xusheng and Lu, Haori and Meng, Yifan and Yang, Fei and Liu, Xialei},
    title     = {Mind the Gap: Preserving and Compensating for the Modality Gap in CLIP-Based Continual Learning},
    booktitle = {Proceedings of the IEEE/CVF International Conference on Computer Vision (ICCV)},
    month     = {October},
    year      = {2025},
    pages     = {3777-3786}
}

@inproceedings{khodabandeh2025lore,
  title     = {{LORE}: Lagrangian-Optimized Robust Embeddings for Visual Encoders},
  author    = {Khodabandeh, Borna and Afzali, Amirabbas and Afsharrad, Amirhossein and Mousavi, Seyed Shahabeddin and Lall, Sanjay and Amini, Sajjad and Moosavi-Dezfooli, Seyed-Mohsen},
  booktitle = {Advances in Neural Information Processing Systems (NeurIPS)},
  year      = {2025}
}

@InProceedings{Liu_2019_CVPR,
author = {Liu, Zihao and Liu, Qi and Liu, Tao and Xu, Nuo and Lin, Xue and Wang, Yanzhi and Wen, Wujie},
title = {Feature Distillation: DNN-Oriented JPEG Compression Against Adversarial Examples},
booktitle = {Proceedings of the IEEE/CVF Conference on Computer Vision and Pattern Recognition (CVPR)},
month = {June},
year = {2019}
}


\end{document}